\documentclass[]{aa}
\usepackage{natbib}
\usepackage[dvips]{graphicx}
\def\ebv{\mbox{$E_{B-V}$\,}}
\def\halpha{\mbox{H$\alpha$}}
\def\hbeta{\mbox{H$\beta$}}
\def\deg{\hbox{$^\circ$}}

\begin{document}

\title{Integral Field Spectroscopy of SN 2002er with PMAS}

\author{L.~Christensen\inst{1}
  \and T.~Becker\inst{1}
  \and K.~Jahnke\inst{1}
  \and A.~Kelz\inst{1} 
  \and M.~M. Roth\inst{1}
  \and S.~F.~S\'anchez\inst{1}
  \and L.~Wisotzki\inst{1,2}
}
\institute{Astrophysikalisches Institut Potsdam, An der Sternwarte 16, 
     14482 Potsdam, Germany
  \and  Potsdam University, Am Neuen Palais 10, 14469 Potsdam, Germany
}
\mail{lchristensen@aip.de}
\date{Received / Accepted}

\abstract{We present observations of the Type Ia supernova \object{SN
2002er} during the brightening phase. The observations were performed
with the Potsdam Multi Aperture Spectrophotometer (PMAS) integral
field instrument. Due to the 8\arcsec$\times$8\arcsec\, field of view
of the spectrograph an accurate background subtraction was possible.
Results from analyses of the evolution of absorption features in
comparisons with other SNe show that SN 2002er is a fairly bright Type
Ia supernova with a peak brightness of M$_B=-19.6\pm0.1$.
}

\maketitle

\section{Introduction}
\label{intro}
Type Ia SNe have been shown not to be the exact standard candles they
were taken for 15 years ago \citep{branch88}. It is now known that
there are several subtypes having slightly different energetics, peak
intensities, light curves, colours, velocities of the outflow, and
synthesized nickel masses. Remarkably, the SNe in elliptical galaxies
are found to have smaller expansion velocities than the SNe in spirals
\citep{hamuy96}. This is not due to differences in the circumstellar
environments of the SNe, but appears to be intrinsic for the
progenitors \citep{lei00}. Correlations between observable properties
have been found, e.g. the light curves of the intrinsically brighter
Type Ia SNe rise and decay more slowly than the faint ones
\citep{phillips93}.

\object{SN 2002er} (17$^h$11$^m$29\fs88  +7\deg59\arcmin44\farcs8,
J2000.0 ) in the galaxy UGC 10743 was discovered on Aug 23 and
reported in IAUC 7959 \citep{iauc7959}. The SN was reported to be a
Type Ia and expected to show optical peak brightness around Sep~6
\citep{iauc7961}. The date of maximum depends on the filter. E.g.
\citet{contardo00} found that the $V$ band peak for 22 SN Ia occurred
from 1 to 3 days after the $B$ band peak. Typically, the time from the
explosion until the peak in the $B$ band light curve is $\sim$20 days
\citep{kirs93,lei00}. The host galaxy has a redshift of
$cz=$2568$\pm7$~km~s$^{-1}$ \citep{up00}, which corresponds to a
distance of 40 Mpc when assuming H$_0$=~65~km~ s$^{-1}$~Mpc$^{-1}$.
The total reddening towards the SN was estimated to be \ebv~=~0.3
\citep{iauc7961} including a Galactic reddening of \ebv~=~0.16
\citep{schlegel98}.

We observed the supernova with the Potsdam Multi Aperture
Spectrophotometer (PMAS) as a target of opportunity on 5 nights in the
brightening phase, with the last observations around the time of peak
intensity. Here we present an analysis of the evolution of some of the
absorption features in the spectra and compare them to other Type Ia
SN.


\section{Observations and data reduction}
\label{data}
Observing with PMAS at the Calar Alto 3.5m telescope, we obtained
spectra of the SN during several dates starting on Aug 31, 2002 and
ending Sep~7. The PMAS instrument has two cameras: A cryogenic
acquisition and guiding camera (A\&G camera) that can be used for
imaging in addition to the integral field spectrograph (IFS)
\citep{pmas00}.

\begin{figure}
\centering \resizebox{\hsize}{!}{\includegraphics{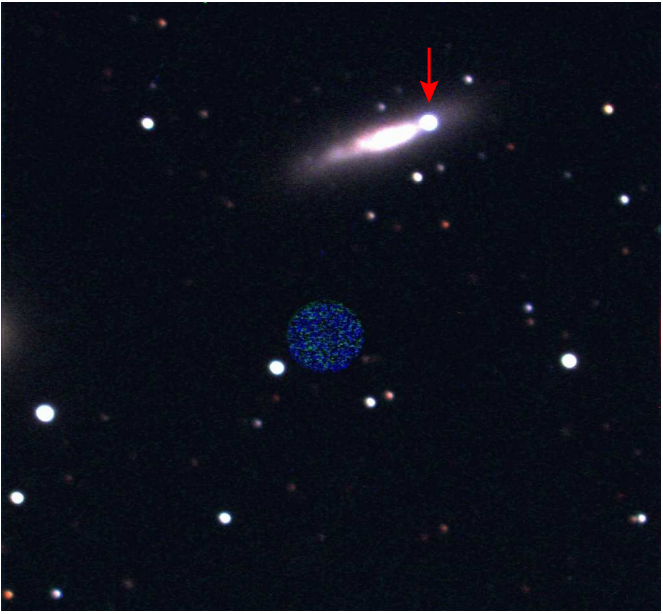}}
\caption{True colour image of the \object{SN 2002er} observed with the
PMAS acquisition and guiding camera. The location of the supernova is
indicated by the arrow. The field of view is 3\farcm4$\times$3\farcm4
and north is up and east is left.  The central green and blue circle
indicates the position of the integral field unit for spectroscopy.
Red, green, and blue correspond to the $R,V$, and $B$ filters, and the
integration times were 60s, 60s, and 120s, respectively.}
\label{fig:sn_ag}
\end{figure}

The A\&G camera has a SITe TK1024 chip with 1k$\times$1k pixels, and a
scale of 0\farcs2 per pixel. The A\&G camera is equipped with optical
filters allowing guiding and imaging in different wavelength
regions. An A\&G image of \object{SN~2002er} can be seen in
Fig~\ref{fig:sn_ag} with a field of view of 3\farcm4$\times$3\farcm4.
The images were reduced with standard procedures, by subtracting a
bias frame and applying flat field correction.

\begin{figure}
\centering \resizebox{\hsize}{!}{\includegraphics{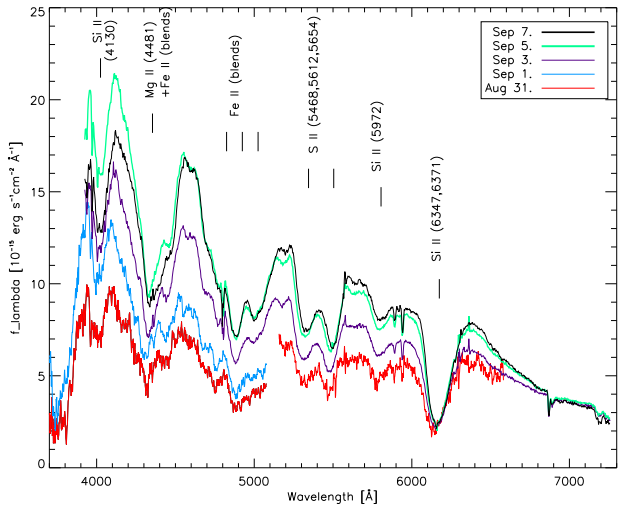}}
\caption{Reduced spectra of the \object{SN 2002er} obtained at 5
dates. The spectra have been corrected for a Galactic extinction of
\ebv = 0.16.  We have identified several absorption lines as
indicated. The Si II~6347, 6371 {\AA} is blue shifted by $\sim12000$
km s$^{-1}$. As can be seen, the SN is observed in the brightening
phase and our observations on Sep~5 and Sep~7 are the brightest. The
spectra obtained with the U600 grating have been smoothed to give the
same spectral resolution as the V300 grating.}
\label{fig:spec}
\end{figure}

The PMAS spectrograph is equipped with 256 fibers coup\-led to a
16$\times$16 lens array. Each fiber has a spatial sampling of
0\farcs5$\times$0\farcs5 on the sky resulting in a field of view of
8\arcsec$\times$8\arcsec. The spectrograph has a SITe ST002A
2k$\times$4k CCD. The 256 spectra have a FWHM of 4 pixels and are
aligned on the CCD with 12 pixels between adjacent spectra making
cross-contamination negligible.

The observations were performed with the gratings used for the primary
science targets during these nights, i.e. a 600 and a 300 gr/mm
grating. The spectral resolution obtained with these gratings is 3 and
6~{\AA}, respectively. Calibration images were obtained following the
science exposures for each night. The calibrations consist of spectra
of emission line lamps (HgNe), and spectra of a continuum lamp
necessary to locate the 256 individual spectra on the CCD, whose
precise positions are shifting as a function of instrument flexure and
grating angle. Observations of the spectrophotometric standard stars
\object{BD~+28\deg 4211}, \object{HD~192281}, and
\object{BD~+25\deg3941} were obtained during the nights, and used for
flux calibrations.  In addition to the SN spectra we have obtained
spectra of the central region of the host galaxy. A summary of the
observations is given in Table~\ref{tab:logfile}.

\begin{table}
\centering
\begin{tabular}{lllll}
\\  \hline \hline  
Date &  UT  & Exptime& Grating & Spec.\ range \\
&  (start)  & (s) & & ({\AA})\\  \hline  \object{SN 2002er}\\ 
Aug 31. & 21:08 & 1320 & U600 &  3470 -- 5070\\
 &        & 600 & U600 & 5140 -- 6650\\ 
Sep 1. & 21:38  & 1920 & U600 &  3470 -- 5070\\ 
Sep 3. & 21:04  & 1500 & V300 & 3925 -- 7260 \\ 
Sep 5. & 20:00  & 900 & V300 & 3925 -- 7260 \\ 
Sep 7. & 20:56 &  900 & V300 & 3925 -- 7260\\
\hline 
UGC 10743\\ 
Sep 3. & 20:23 & 1200 & V300 &  3925 -- 7260\\
\hline
\end{tabular}
\caption[]{ Log of the observations with the spectrograph.}
\label{tab:logfile}
\end{table}

\subsection{Reduction of spectra}
The data extraction and reduction of the spectra was done with
P3D\_online\footnote{Further information can be found at the PMAS
webpages http://www.aip.de/groups/opti/pmas/OptI\_pmas.html}, an IDL
based software package created for reduction of PMAS data
\citep{becker01}. The raw CCD frames were bias subtracted, and the
extracted spectra are flat-fielded correcting for throughput
variations of individual fibers using twilight sky exposures. The flux
calibration was performed with standard procedures in IRAF by
comparing the extracted spectra of the standard star with table
values, and using table values from Calar Alto in order to correct for
the atmospheric extinction. With exposures of Hg and Ne lamps obtained
right after the science exposures, the spectra were wavelength
calibrated.

The reduced spectra were contained in a 16$\times$16$\times$1024
pixels data cube (with 2$\times$2 binning). The average background was
estimated from the 60 spectra at the edges of the field of view.  We
first took an average of all 256 spectra and subtracted the average
background. This produces a noisy spectrum of the supernova, but no
flux is lost due to any chosen aperture. A higher S/N spectrum of the
supernova was obtained by co-adding the spectra within a radius of
1\farcs5 of the center of the SN, and then scaling to the level of the
low S/N ``wide aperture'' spectrum.

We did not have reliable spectrophotometric standard star observations
on Aug 31 and Sep 1. In order to perform accurate flux calibration we
analysed the $V$ band A\&G images from all dates. A model of the host
was subtracted from the images and the uncontaminated flux from the SN
was found. With relative photometry between the SN and field stars we
are able to rescale the flux of the red part of the spectrum on Aug
31, similar to the technique described in \cite{barwig87}. We did not
have appropriate A\&G data to re-calibrate the blue part of the
spectra using a similar method. We estimate that the flux-levels for
the blue spectra on Aug 31 is uncertain by $\sim$20\% estimated from
comparison with the flux for the red spectrum. The flux of the blue
spectrum at Sep 1 should lie between the Aug 31 and Sep 3 spectral
flux, so we estimate that this flux is uncertain by $\sim$15\%. The
calibrated spectra are shown in Fig.~\ref{fig:spec}.

\section{Spectral evolution}
\label{evol}
We have identified several absorption lines which clearly confirm the
supernova as a Type Ia. During the observations the supernova
increased its brightness, and the spectrum from Sep~5 is the brightest
one in the blue region, while the red end is brightest at Sep~7 in
accordance with the change of peak intensity with wavelength in
\citet{contardo00}.

Of particular interest is the prominent Si II~6347, 6371~{\AA} (with a
$g\!f$ average of 6355~{\AA}) absorption feature near the maximum of a
Type Ia SN. The P~Cygni profile of this line indicates large expansion
velocities, which can be estimated from position of the blue edge of
the absorption profile. As the photosphere expands, the absorption
feature moves towards redder wavelengths as can be seen in
Table~\ref{tab:spec}. Correcting for the redshift of the host galaxy,
the velocity evolution determined from the center of the Si II line
changes from 12300 km s$^{-1}$ on Aug 31 to 11200 km s$^{-1}$ on
Sep~7. Considering the very large range of expansion velocities of
other Type Ia SNe of 10000--15000 km s$^{-1}$ near the maximum
\citep{branch88}, the expansion velocity of \object{SN 2002er} is
normal for a bright SN. The blue shift of the Si II line could,
however, also be caused by increasing metallicities of the progenitor
\citep{lentz00}.

Different lines do not have identical velocities. We see for example
that the Mg II line has larger velocities than Si II by roughly 2000
km s$^{-1}$, which suggests a layered structure of the explosion
products \citep{patat96}.

We have estimated the $B$ band magnitude by convolving the spectra
with a transmission curve of the Bessell $B$ filter. The peak
intensity in the $B$ band occurred between our last two spectra,
i.e. on Sep~6, confirming the prediction of \citet{iauc7961}. At
redder wavelengths the peak occurred later, see Table~\ref{tab:spec}.
 
\begin{table}
\centering
\begin{tabular}{llllll}
\\  
\hline \hline  date&   min. $\lambda$ & $W$ & expansion  & $B$ &
$V$\\ 
& ({\AA})& ({\AA}) & (km s$^{-1}$) & & \\ 
\hline  
Aug 31. & 6148 & 110 & 12300& -- & 15.52\\ 
Sep 3.  & 6159  & 118 & 11800& 15.05 &14.83\\    
Sep 5. &  6166  & 122 & 11500& 14.76 & 14.61\\    
Sep 7.& 6172  & 122 & 11200& 14.82 & 14.57\\    
\hline
\end{tabular}
\caption[]{Evolution of the Si II 6355~{\AA} feature with time. The
$B$ and $V$ band magnitudes are calculated by convolving the spectra
with the transmission of Bessell $B$ and $V$ filters,
respectively. The $B$ band magnitude on Aug 31 is not given since the
spectrum at this wavelength is not properly flux calibrated.}
\label{tab:spec}
\end{table}

The three ``notches'' at 4550, 4650 and 5150~{\AA} are visible in
several Type Ia SNe spectra \citep{filip97}. In Fig.~\ref{fig:spec}
these features become more pronounced with time as transitions in Fe
become more important. The two S II lines at 5300 and 5500~{\AA} are
almost of equal strength, which indicates that this SN was a bright
event such as \object{SN 1981B}, however not as bright at the very
luminous \object{SN~1991T}. In the fainter SNe the red line is much
stronger than the blue \citep{nugent95}.

The ratio of the Si II absorption at 5972~{\AA} to the 6355~{\AA}
feature increases with decreasing luminosity \citep{nugent95}. This
ratio gives $\mathcal R$(Si~II)$=0.21~\pm~0.01$ for
\object{SN~2002er}, a value typical for the most luminous Type Ia
events. We do not see a significant evolution of this ratio with time.

The two narrow absorption features at $\sim$5900~{\AA} have been
identified as interstellar Na 5890~{\AA} in the Milky Way and the host
galaxy respectively.

\subsection{UGC 10743: The host galaxy}
\begin{figure}
\centering \resizebox{\hsize}{!}{\includegraphics{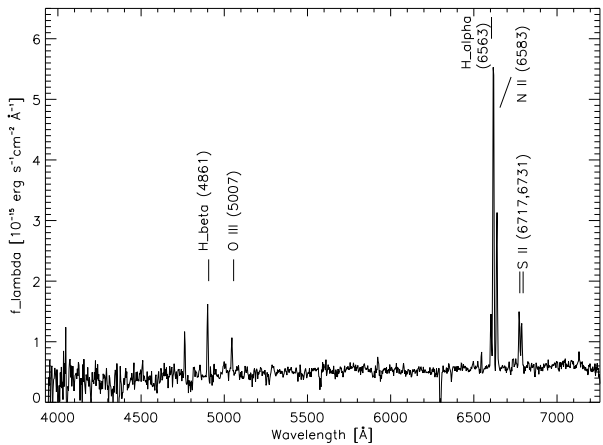}}
\caption{Reduced spectrum of the central region of \object{UGC
10743}. The spectrum is flat and dominated by emission lines.}
\label{fig:hostspec}
\end{figure}

The very flat spectrum of the host clearly shows \halpha, \hbeta,
[O~III] 5007~{\AA}, [S~II] 6717, 6731~{\AA} and [N~II] 6583~{\AA}
emission lines redshifted by 2560$\pm$20 km s$^{-1}$, see
Fig.~\ref{fig:hostspec}. These lines and the [O I] sky emission lines
are subtracted from the spectra of the supernova. We have analysed the
surface brightness distribution of the host on the A\&G images, and
detected no significant contribution from a bulge classifying the
galaxy as an Sc type. This is in agreement with a flat spectrum,
characteristic for late type galaxies, supporting that luminous SNe Ia
are found in these galaxies \citep{hamuy96}.

\section{Discussion and outlook}
\label{disc}
We estimate the absolute magnitude in the $B$ band M$_B=-19.0\pm 0.1$
at the peak of the light curve taking into account the Galactic
extinction of \ebv=~0.16 and using the Milky Way extinction curve from
\citet{seaton79}. Including the extinction in the host the peak is at
M$_B=-19.6\pm0.1$. The average brightness of SNe Ia is M$_B=-19.3$
\citep{hille00,lei01}.  The dispersion at maximum is $\sigma_M \sim
0.3$ \citep{hamuy96}. Thus SN 2002er was a fairly bright SN.

Together with the measurements of the outflow velo\-city, the
absorption line strengths and the $\mathcal R$(Si II) ratio suggest
that \object{SN 2002er} was a typical bright Type Ia SN, similar to
other intrinsically bright SNe such as \object{SN 1981B}. This is also
in agreement with the indication that brighter SNe are preferably
found in late type galaxies, suggesting a range in progenitor
properties \citep{hamuy96}. Further modeling is needed to determine
abundances and late time spectra are necessary for calculating the
bolometric luminosity as well as the explosively synthesized Ni mass.

With an integral field spectrograph one can get an accurate background
estimate for later subtraction from the object spectrum. Specifically,
removal of the contaminating host galaxy is better performed than with
classical slit spectroscopy. The present observations make \object{SN
2002er} the second SN ever observed with an integral field unit. The
first was the serendipitous discovery of \object{SN~1998cf} in
\object{NGC 3504} as reported by \citet{emsel98} and \citet{garn98}.

The standard candle nature of Type Ia SNe has been used to infer a
non-zero cosmological constant \citep{riess98,perlmutter99}.
Determining accurate spectrophotometric evolution of several subtypes
of SNe Ia is essential for calculating the absolute peak magnitude,
and thereby the distance to the cosmological SNe. One of the key
problems is the correction of the flux by absorption of dust in the
host galaxy. Detailed observations of nearby SNe with a dedicated
integral field instrument, SNIFS \citep{alder02}, will allow to
disentangle the dust contents in the host from the intrinsic colour of
the SNe, and then calibrate the more distant SNe. The future SNAP
satellite \citep{deusta00}, will by means of detecting distant Type Ia
SNe, determine cosmological parameters with a small uncertainty. Both
projects will include IFSs in order to make use of the possibility for
spatially resolved and simultaneous spectroscopy of the supernovae and
their environments.  

Our observations showed that PMAS is an efficient instrument for doing
spectrophotometry of a point source especially with a high background
contamination, and without slit losses. It was also demonstrated that
the A\&G camera can be used to perform differential photometry with
field stars to increase the absolute spectrophotometric calibration
under non-photometric conditions.

\begin{acknowledgements}
L.~Christensen acknowledges support by the German Verbundforschung
associated with the ULTROS project, grant
no. 05AE2BAA/4. S.F.~S\'anchez acknowledges the support from the
Euro3D Research Training Network, grant
no. HPRN-CT2002-00305. K. Jahnke and L. Wisotzki acknowledge a DFG
travel grant under Wi~1369/12-1.
\end{acknowledgements}

\bibliography{sn2002er}

\end{document}